\documentclass{llncs}
\usepackage{amsmath}
\usepackage{amsfonts,amssymb}
\usepackage[thinlines,thiklines]{easybmat}
\usepackage{multirow}
\usepackage{url}
\usepackage{colortbl}

\newcommand{\Z}{\mathbb{Z}} % les entiers relatifs

\begin{document}

\mainmatter              % start of the contributions
\title{Security Estimates for Quadratic Field Based Cryptosystems}
\titlerunning{Security Estimates for Quadratic Field Based Cryptosystems}  % abbreviated title (for running head)
%                                     also used for the TOC unless
%                                     \toctitle is used
%
\author{Jean-Fran\c{c}ois Biasse\inst{1} \and Michael J. Jacobson,
  Jr.\inst{2}\thanks{The second author
  is supported in part by NSERC of Canada.} \and Alan K. Silvester\inst{3}}
\authorrunning{J-F Biasse,M. J. Jacobson, Jr., and A. K. Silvester}   % abbreviated author list (for running head)
%
%%%% list of authors for the TOC (use if author list has to be modified)
\tocauthor{Jean-Fran\c{c}ois Biasse  Michael J. Jacobson, Jr. Alan K. Silvester}
\institute{ \'{E}cole Polytechnique, 91128 Palaiseau, France\\
\email{biasse@lix.polytechnique.fr}
\and
Department of Computer Science,
University of Calgary \\
2500 University Drive NW, Calgary, Alberta, Canada T2N 1N4\\
\email{jacobs@cpsc.ucalgary.ca}
\and
Department of Mathematics and Statistics,
University of Calgary \\
2500 University Drive NW, Calgary, Alberta, Canada T2N 1N4\\
\email{aksilves@math.ucalgary.ca}
}

\maketitle              % typeset the title of the contribution

\begin{abstract}
We describe implementations for solving the discrete logarithm problem
in the class group of an imaginary quadratic field and in the
infrastructure of a real quadratic field.  The algorithms used
incorporate improvements over previously-used algorithms, and
extensive numerical results are presented demonstrating their
efficiency.  This data is used as the basis for extrapolations, used
to provide recommendations for parameter sizes providing approximately
the same level of security as block ciphers with $80,$ $112,$ $128,$
$192,$ and $256$-bit symmetric keys.
\end{abstract}

\section{Introduction}

Quadratic fields were proposed as a setting for public-key
cryptosystems in the late 1980s by Buchmann and Williams
\cite{BWKeyEx,BWKeyExReal}.  There are two types of quadratic fields,
imaginary and real.  In the imaginary case, cryptosystems are based on
arithmetic in the ideal class group (a finite abelian group), and the
discrete logarithm problem is the computational problem on which the
security is based.  In the real case, the so-called infrastructure is
used instead, and the security is based on the analogue of the
discrete logarithm problem in this structure, namely the principal ideal
problem.

Although neither of these problems is resistant to quantum computers,
cryptography in quadratic fields is nevertheless an interesting
alternative to more widely-used settings.  Both discrete logarithm
problems can be solved in subexponential time using index calculus
algorithms, but with asymptotically slower complexity than the
state-of-the art algorithms for integer factorization and computing
discrete logarithms in finite fields.  In addition, the only known
relationship to the quadratic field discrete logarithm problems from
other computational problems used in cryptography is that integer
factorization reduces to both of the quadratic field problems.  Thus,
both of these are at least as hard as factoring, and the lack of known
relationships to other computational problems
implies that the breaking of other cryptosystems, such as those based
on elliptic or hyperelliptic curves, will not necessarily break those
set in quadratic fields.  Examining the security of quadratic field
based cryptosystems is therefore of interest.

The fastest algorithms for solving discrete logarithm problem in
quadratic fields are based on an improved version of Buchmann's
index-calculus algorithm due to Jacobson \cite{Jdl}.  The algorithms
include a number of practical enhancements to the original algorithm
of Buchmann \cite{BSub}, including the use of self-initialized sieving
to generate relations, a single large prime variant, and
practice-oriented algorithms for the required linear algebra.  These
algorithms enabled the computation of a discrete logarithm in the
class group of an imaginary quadratic field with $90$ decimal digit
discriminant \cite{HJWnonint_full}, and the solution of the principal
ideal problem for a real quadratic field with $65$ decimal digit
discriminant \cite{JSWkeyex}.

Since this work, a number of further improvements have been proposed.
Biasse \cite{biasse} presented practical improvements to the
corresponding algorithm for imaginary quadratic fields, including a
double large prime variant and improved algorithms for the required
linear algebra.  The resulting algorithm was indeed faster then the
previous state-of-the-art and enabled the computation of the ideal
class group of an imaginary quadratic field with $110$ decimal digit
discriminant.  These improvements were adapted to the case of real
quadratic fields by Biasse and Jacobson \cite{BJregulator}, along with
the incorporation of a batch smoothness test of Bernstein
\cite{bernstein}, resulting in similar speed-ups in that case.

In this paper, we adapt the improvements of Biasse and Jacobson to the
computation of discrete logarithms in the class group of an imaginary
quadratic field and the principal ideal problem in the infrastructure
of a real quadratic field.  We use versions of the algorithms that
rely on easier linear algebra problems than those described in
\cite{Jdl}.  In the imaginary case, this idea is due to Vollmer
\cite{Vdl}; our work represents the first implementation of his
method.  Our data obtained shows that our algorithms are indeed faster
than previous methods.  We use our data to estimate parameter sizes
for quadratic field cryptosystems that offer security equivalent to
NIST's five recommended security levels \cite{NIST-keys}.  In the
imaginary case, these recommendations update previous results of Hamdy
and M{\"o}ller \cite{HMiq}, and in the real case this is the first
time such recommendations have been provided.

The paper is organized as follows.  In the next section, we briefly
recall the required background of ideal arithmetic in quadratic
fields, and give an overview of the index-calculus algorithms for
solving the two discrete logarithms in Section~\ref{sec:dlp}.  Our
numerical results are described in Section~\ref{sec:results}, followed
by the security parameter estimates in Section~\ref{sec:estimates}.

\section{Arithmetic in Quadratic Fields}

We begin with a brief overview of arithmetic in quadratic fields.  For
more details on the theory, algorithms, and cryptographic applications
of quadratic fields, see \cite{JWPellBook}.

Let $K = \mathbb{Q}(\sqrt{\Delta})$ be the quadratic field of
discriminant $\Delta,$ where $\Delta$ is a non-zero integer congruent
to $0$ or $1$ modulo $4$ with $\Delta$ or $\Delta/4$ square-free.  The
integral closure of $\mathbb{Z}$ in $K$, called the maximal order, is
denoted by $\mathcal{O}_{\Delta}.$ The ideals of $\mathcal{O}_{\Delta}$ are
the main objects of interest in terms of cryptographic applications.
An ideal can be represented by the two dimensional $\mathbb{Z}$-module
\begin{equation*}
\mathfrak{a} = s \left[a \mathbb{Z} + \frac{b +
    \sqrt{\Delta}}{2} \mathbb{Z} \right]\;,
\end{equation*}
where $a,b,s \in \Z$ and $4a ~|~ b^2 - \Delta.$ The integers $a$ and
$s$ are unique, and $b$ is defined modulo $2a.$ The ideal
$\mathfrak{a}$ is said to be primitive if $s = 1.$ The norm of
$\mathfrak{a}$ is given by $\mathcal{N}(\mathfrak{a}) = a s^2.$

Ideals can be multiplied using Gauss' composition formulas for
integral binary quadratic forms.  Ideal norm respects this operation.
The prime ideals of $\mathcal{O}_{\Delta}$ have the form $p \mathbb{Z}
+ (b_p + \sqrt{\Delta})/2 \mathbb{Z}$ where $p$ is a prime that is
split or ramified in $K,$ i.e., the Kronecker symbol $(\Delta/p) \neq
-1.$ As $\mathcal{O}_{\Delta}$ is a Dedekind domain, every ideal can
be factored uniquely as a product of prime ideals.  To factor
$\mathfrak{a},$ it suffices to factor $\mathcal{N}(\mathfrak{a})$ and,
for each prime $p$ dividing the norm, determine whether the prime
ideal $\mathfrak{p}$ or $\mathfrak{p}^{-1}$ divides $\mathfrak{a}$
according to whether $b$ is congruent to $b_p$ or $-b_p$ modulo $2p.$

Two ideals $\mathfrak{a}, \mathfrak{b}$ are said to be equivalent,
denoted by $\mathfrak{a} \sim \mathfrak{b},$ if there exist $\alpha,
\beta \in \mathcal{O}_{\Delta}$ such that $(\alpha) \mathfrak{a} =
(\beta) \mathfrak{b},$ where $(\alpha)$ denotes the principal ideal
generated by $\alpha.$ This is in fact an equivalence relation, and
the set of equivalence classes forms a finite abelian group called the
class group, denoted by $Cl_\Delta.$ Its order is called the class
number, and is denoted by $h_\Delta.$

Arithmetic in the class group is performed on reduced ideal
representatives of the equivalence classes.  An ideal $\mathfrak{a}$
is reduced if it is primitive and $\mathcal{N}(\mathfrak{a})$ is a minimum
in $\mathfrak{a}.$ Reduced ideals have the property that $a,b <
\sqrt{|\Delta|},$ yielding reasonably small representatives of each
group element.  The group operation then consists of multiplying two
reduced ideals and computing a reduced ideal equivalent to the
product.  This operation is efficient and can be performed in
$O(\log^2 |\Delta|)$ bit operations.

In the case of imaginary quadratic fields, we have $h_\Delta
\approx \sqrt{|\Delta|},$ and that every element in $Cl_\Delta$ contains
exactly one reduced ideal. Thus, the ideal class group can be used as
the basis of most public-key cryptosystems that require arithmetic in
a finite abelian group.  The only wrinkle is that computing the class
number $h_\Delta$ seems to be as hard as solving the discrete
logarithm problem, so only cryptosystems for which the group order is
not known can be used.

In real quadratic fields, the class group tends to be small; in fact,
a conjecture of Gauss predicts that $h_\Delta = 1$ infinitely often,
and the Cohen-Lenstra heuristics \cite{CLheur} predict that this
happens about $75 \%$ of the time for prime discriminants. Thus, the
discrete logarithm problem in the class group is not in general
suitable for cryptographic use.  

Another consequence of small class groups in the real case is that
there are no longer unique reduced ideal representatives in each
equivalence class.  Instead, we have that $h_\Delta R_\Delta \approx
\sqrt{\Delta},$ where the regulator $R_\Delta$ roughly approximates
how many reduced ideals are in each equivalence class.  Thus, since
$h_\Delta$ is frequently small, there are roughly $\sqrt{\Delta}$
equivalent reduced ideals in each equivalence class.  The
infrastructure, namely the set of reduced principal ideals, is used
for cryptographic purposes instead of the class group.  Although this
structure is not a finite abelian group, the analogue of
exponentiation (computing a reduced principal ideal $(\alpha)$ with
$\log \alpha$ as close to a given number as possible) is efficient and
can be used as a one-way problem suitable for public-key cryptography.
The inverse of this problem, computing an approximation of the unknown
$\log \alpha$ from a reduced principal ideal given in $\Z$-basis
representation, is called the principal ideal problem or
infrastructure discrete logarithm problem, and is believed to be of
similar difficulty to the discrete logarithm problem in the class
group of an imaginary quadratic field.

\section{Solving The Discrete Logarithm Problems}\label{sec:dlp}

The fastest algorithms in practice for computing discrete logarithms
in the class group and infrastructure use the index-calculus
framework.  Like other index-calculus algorithms, these algorithms
rely on finding certain smooth quantities, those whose prime divisors
are all small in some sense.  In the case of quadratic fields, one
searches for smooth principal ideals for which all prime ideal
divisors have norm less than a given bound $B.$ The set of prime
ideals $\mathfrak{p}_1, \dots, \mathfrak{p}_n$ with
$\mathcal{N}(\mathfrak{p}_i) \leq B$ is called the factor base,
denoted by $\mathcal{B}.$

A principal ideal $(\alpha) = \mathfrak{p}_1^{e_1} \cdots
\mathfrak{p}_n^{e_n}$ with $\alpha \in K$ that factors completely over
the factor base yields the relation $(e_1,\dots,e_n, \log |\alpha|).$
In the imaginary case, the $\log |\alpha|$ coefficients are not
required and are ignored.  The key to the index-calculus approach is
the fact, proved by Buchmann \cite{BSub}, that the set of all
relations forms a sublattice $\Lambda \subset \mathbb{Z}^n \times
\mathbb{R}$ of determinant $h_\Delta R_\Delta$ as long as the prime
ideals in the factor base generate $Cl_\Delta.$ This follows, in part,
due to the fact that $L,$ the integer component of $\Lambda,$ is the
kernel of the homomorphism $\phi:\mathbb{Z}^n \mapsto Cl_\Delta$ given
by $\mathfrak{p}_1^{e_1} \cdots \mathfrak{p}_n^{e_n}$ for
$(e_1,\dots,e_n) \in \mathbb{Z}^n.$ The homomorphism theorem then
implies that $\mathbb{Z}^n / L \cong Cl_\Delta.$ In the imaginary
case, where the $\log |\alpha|$ terms are omitted, the relation
lattice consists only of the integer part, and the corresponding
results were proved by Hafner and McCurley \cite{HMSub}.

The main idea behind the algorithms described in \cite{Jdl} for
solving the class group and infrastructure discrete logarithm problems
is to find random relations until they generate the entire relation
lattice $\Lambda.$ Suppose $A$ is a matrix whose rows contain the
integer coordinates of the relations, and $\vec{v}$ is a vector
containing the real parts.  To check whether the relations generate
$\Lambda,$ we begin by computing the Hermite normal form of $A$ and
then calculating its determinant, giving us a multiple $h$ of the
class number $h_\Delta.$ We also compute a multiple of the regulator
$R_\Delta.$ Using the analytic class number formula and Bach's
$L(1,\chi)$-approximation method \cite{BBounds}, we construct bounds
such that $h_\Delta R_\Delta$ itself is the only integer multiple of
the product of the class number and regulator satisfying $h^* <
h_\Delta < 2h^*$; if $h R$ satisfies these bounds, then $h$ and $R$
are the correct class number and regulator and the set of relations
given in $A$ generates $\Lambda.$

A multiple $R$ of the regulator $R_\Delta$ can be computed either from
a basis of the kernel of the row-space of $A$ (as in \cite{Jdl}) or by
randomly sampling from the kernel as described by Vollmer \cite{Vreg}.
Every kernel vector $\vec{x}$ corresponds to a multiple of the
regulator via $\vec{x} \cdot \vec{v} = m R_\Delta.$ Given $\vec{v}$
and a set of kernel vectors, an algorithm of Maurer
\cite[Sec~12.1]{MMaurer} is used to compute the ``real GCD'' of the
regulator multiples with guaranteed numerical accuracy, where the real
GCD of $m_1 R_\Delta$ and $m_2 R_\Delta$ is defined to be
$\gcd(m_1,m_2) R_\Delta.$

To solve the discrete logarithm problem in $Cl_\Delta,$ we compute the
structure of $Cl_\Delta,$ i.e., integers $m_1, \dots, m_k$ with
$m_{i+1} ~|~ m_i$ for $i=1,\dots,k-1$ such that $Cl_\Delta \cong \Z /
m_1 \Z \times \cdots \times \Z / m_k \Z,$ and an explicit isomorphism
from $\Z^n$ to $\Z / m_1 \Z \times \cdots \times \Z / m_k.$ Then, to
compute $x$ such that $\mathfrak{g}^x \sim \mathfrak{a},$ we find
ideals equivalent to $\mathfrak{g}$ and $\mathfrak{a}$ that factor
over the factor base and maps these vectors in $\Z^n$ to $\Z / m_1 \Z
\times \cdots \times \Z / m_k,$ where the discrete logarithm problem
can be solved easily.

To solve the infrastructure discrete logarithm problem for
$\mathfrak{a},$ we find an ideal equivalent to $\mathfrak{a}$ that
factors over the factor base.  Suppose the factorization is given by
$\vec{v} \in \Z^n.$ Then, since $L$ is the kernel of $\phi,$ if
$\mathfrak{a}$ is principal, $\vec{v}$ must be a linear combination of
the elements of $L.$ This can be determined by solving $\vec{x} A =
\vec{v},$ where as before the rows of $A$ are the vectors in $L.$
Furthermore, we have $\log \alpha = \vec{x} \cdot \vec{v}
\pmod{R_\Delta}$ is a solution to the infrastructure discrete
logarithm problem.  The approximation of $\log \alpha$ is computed to
guaranteed numerical accuracy using another algorithm of Maurer
\cite[Sec~5.5]{MMaurer}.

If it is necessary to verify the solvability of the problem instance,
then one must verify that the relations generate all of $\Lambda,$ for
example, as described above.  The best methods for this certification
are conditional on the Generalized Riemann Hypothesis, both for their
expected running time and their correctness.  However, in a
cryptographic application, it can safely be assumed that the problem
instance does have a solution (for example, if it comes from the
Diffie-Hellman key exchange protocol), and simplifications are
possible.  In particular, the correctness of the computed solution can
be determined without certifying that the relations generate
$\Lambda,$ for example, by verifying that $\mathfrak{g}^x =
\mathfrak{a}.$ As a result, the relatively expensive linear algebra
required (computing Hermite normal form and kernel of the row space)
can be replaced by linear system solving.

In the imaginary case, if the discrete logarithm is known to exist,
one can use an algorithm due to Vollmer
\cite{Vdl,vollmer-thesis:2003}.  Instead of computing the structure of
$Cl_\Delta,$ one finds ideals equivalent to $\mathfrak{g}$ and
$\mathfrak{a}$ that factor over the factor base.  Then, combining
these factorizations with the rest of the relations and solving a
linear system yields a solution of the discrete logarithm problem.  If
the linear system cannot be solved, then the relations do not generate
$\Lambda,$ and the process is simply repeated after generating some
additional relations.  The expected asymptotic complexity of this
method, under reasonable assumptions about the generation of
relations, is $O(L_{|\Delta|}[1/2,3 \sqrt{2}/4 + o(1)])$
\cite{vollmer-thesis:2003,BVforms}, where
\begin{equation*}
L_N[e,c] = \exp\big(c\,(\log N)^e (\log \log N)^{1-e}\big)
\end{equation*}
for $e,c$ constants and $0 \leq e \leq 1.$ In practice, all the
improvements to relation generation and simplifying the relation
matrix described in \cite{biasse} can be applied.  When using
practical versions for generating relations, such as sieving as
described in \cite{Jdl}, it is conjectured that the algorithm has
complexity $O(L_{|\Delta|}[1/2,1+o(1)]).$

In the real case, we also do not need to compute the Hermite normal
form, as only a multiple of $R_\Delta$ suffices. The consequence of
not certifying that we have the true regulator is that the solutions
obtained for the infrastructure discrete logarithm problem may not be
minimal.  However, for cryptographic purposes this is sufficient, as
these values can still be used to break the corresponding protocols in
the same way that a non-minimal solution to the discrete logarithm
problem suffices to break group-based protocols.  Thus, we use
Vollmer's approach \cite{Vreg} based on randomly sampling from the
kernel of $A.$ This method computes a multiple that is with high
probability equal to the regulator in time $O(L_{|\Delta|}[1/2,3
\sqrt{2}/4 + o(1)])$ by computing the multiple corresponding to random
elements in the kernel of the row space of $A.$ These random elements
can also be found by linear system solving.  The resulting algorithm
has the same complexity as that in the imaginary case.  In practice,
all the improvements described in \cite{BJregulator} can be applied.
When these are used, including sieving as described in \cite{Jdl}, we
also conjecture that the algorithm has complexity
$O(L_{|\Delta|}[1/2,1+o(1)]).$

\section{Implementation and Numerical Results}\label{sec:results}

Our implementation takes advantage of the latest practical
improvements in ideal class group computation and regulator
computation for quadratic number fields, described in detail in
\cite{biasse,BJregulator}.  In the following, we give a brief outline
of the methods we used for the experiments described in this paper.

To speed up the relation collection phase, we combined the double
large prime variation with the self-initialized quadratic sieve
strategy of \cite{Jdl}, as descried in \cite{biasse}.
%The double
%large prime variation consists of defining a bound $B' > B$, and
%allowing relations of the form $(\alpha) =
%\mathfrak{p}_1^{e_1}\hdots\mathfrak{p}_n^{e_n}\mathfrak{p}\mathfrak{p}',$
%where $\mathfrak{p}_i\in\mathcal{B}$ for $i\leq n$ and
%$\mathcal{N}(\mathfrak{p}),\mathcal{N}(\mathfrak{p}')\leq B'$.
This results in a considerable speed-up in the time required for
finding a relation, at the cost of a growth of the dimensions of the
relation matrix. We also used Bernstein's batch smoothness test
\cite{bernstein} to enhance the relation collection phase as described
in \cite{BJregulator}, by simultaneously testing residues produced by
the sieve for smoothness.
%During
%sieving, a large amount of integer residues are tested for
%smoothness. In \cite{Jdl,biasse}, this was done by simple trial
%division. Bernstein's method can be used to simultaneously test the
%smoothness of a given set of residues. 
%This algorithm was used for the
%relation collection in the real case by Biasse and Jacobson
%\cite{BJregulator}, and we applied it for our experiments in both
%imaginary and real case.

The algorithms involved in the linear algebra phase are highly
sensitive to the dimensions of the relation matrix. As the double
large prime variation induces significant growth in the dimensions of
the relation matrix, one needs to perform Gaussian elimination to
reduce the number of columns in order to make the linear algebra phase
feasible. We used a graph-based elimination strategy first described
by Cavallar \cite{cavallar} for factorization, and then adapted by
Biasse \cite{biasse} to the context of quadratic fields. At the end of
the process, we test if the resulting matrix $A_{red}$ has full rank
by reducing it modulo a word-sized prime. If not, we collect more
relation and repeat the algorithm.

For solving the discrete logarithm problem in the imaginary case, we
implemented the algorithm due to Vollmer
\cite{Vdl,vollmer-thesis:2003} . Given two ideals $\mathfrak{a}$ and
$\mathfrak{g}$ such that $\mathfrak{g}^x \sim \mathfrak{a}$ for some
integer $x$, we find two extra relations $(e_1,\dots,e_n,1,0)$ and
$(f_1,\dots,f_n,0,1)$ such that
$\mathfrak{p}_1^{e_1}\cdots\mathfrak{p}_n^{e_n} \mathfrak{g} \sim (1)$
and $\mathfrak{p}_1^{f_1}\cdots\mathfrak{p}_n^{f_n} \mathfrak{a}^{-1}
\sim (1)$ over the extended factor base $\mathcal{B}\cup\left\lbrace
\mathfrak{g},\mathfrak{a}^{-1}\right\rbrace $. The extra relations are
obtained by multiplying $\mathfrak{a}^{-1}$ and $\mathfrak{g}$ by
random power products of primes in $\mathcal{B}$ and sieving with the
resulting ideal to find an equivalent ideal that is smooth over
$\mathcal{B}.$ Once these relations have been found, we construct the
matrix
\[ A' := \left( 
   \begin{BMAT}(@)[2pt,1.5cm,1.5cm]{c.c}{c.c}
   \begin{BMAT}(2pt,1cm,1cm){c}{c}
A
  \end{BMAT} &
\begin{BMAT}(2pt,0.5cm,1cm){c}{c}
(0)
  \end{BMAT} \\
\begin{BMAT}(2pt,1cm,0.5cm){ccc}{cc} 
	e_1 & \dots & e_n \\
	f_1 & \dots & f_n
\end{BMAT} &
\begin{BMAT}(2pt,0.5cm,0.5cm){cc}{cc} 
	1 & 0 \\
	0 & 1
\end{BMAT} 
\end{BMAT}
   \right),
\]
and solve the system $\vec{x} A' = (0,\hdots,0,1)$. The last
coordinate of $\vec{x}$ necessarily equals the discrete logarithm $x$.
We used \texttt{certSolveRedLong} from the IML library \cite{iml} to
solve these linear systems.

As the impact of Vollmer's and Bernstein's algorithms on the overall
time for class group and discrete logarithm computation in the
imaginary case had not been studied, we provide numerical data in
Table~\ref{DL_CL:tab} for discriminants of size between 140 and 220
bits. The timings, given in seconds, are averages of three different
random prime discriminants, obtained with 2.4 GHz Opterons with 8GB or
memory. We denote by ``DL'' the discrete logarithm computation using
Vollmer's method and by ``CL'' the class group computation.  ``CL
Batch'' and ``DL Batch'' denote the times obtained when also using
Bernstein's algorithm.  We list the optimal factor base size for each
algorithm and discriminant size (obtained via additional numerical
experiments), the time for each of the main parts of the algorithm,
and the total time.  In all cases we allowed two large primes and took
enough relations to ensure that $A_{red}$ have full rank.  Our results
show that enhancing relation generation with Bernstein's algorithm is
beneficial in all cases.  In addition, using Vollmer's algorithm for
computing discrete logarithms is faster than the approach of
\cite{Jdl} that also requires the class group.
\begin{table}[!t]
\caption{\label{DL_CL:tab}Comparison between class group computation and Vollmer Algorithm}
\renewcommand{\arraystretch}{1.1}
%\small 
\begin{center}
 \begin{tabular}{|r|r|r|r|r|r|r|}
\hline
\multicolumn{1}{|>{\ }c<{\ }|}{Size} & 
\multicolumn{1}{>{\ }c<{\ }|}{Strategy} & 
\multicolumn{1}{>{\ \ }c<{\ \ }|}{$|\mathcal{B}|$} &
\multicolumn{1}{>{\ }c<{\ }|}{Sieving} & 
\multicolumn{1}{>{\ }c<{\ }|}{Elimination} & 
\multicolumn{1}{>{\ }c<{\ }|}{Linear algebra} & 
\multicolumn{1}{>{\ \ }c<{\ \ }|}{Total} \\
\hline
 \multirow{4}{*}{140} & CL &  200 & 2.66 & 0.63 & 1.79 & 5.08\\
  & CL Batch &  200  & 1.93 & 0.65 & 1.78 & 4.36\\
  & DL  &  200 & 2.57 & 0.44 &  0.8 & 3.81 \\ 
  & DL batch  & 200 & 1.92 & 0.41 &  0.76 & 3.09 \\
\hline
 \multirow{4}{*}{160} & CL  &  300 & 11.77 & 1.04 & 8.20 & 21.01\\
  & CL Batch &  300  & 9.91 & 0.87 & 8.19 & 18.97\\
  & DL  &  350 & 10.17 & 0.73  & 2.75 & 13.65 \\
  & DL batch  &  400 & 6.80 & 0.96   & 3.05 & 10.81 \\
\hline
 \multirow{4}{*}{180} & CL  &  400  & 17.47 & 0.98 & 12.83 & 31.28\\
  & CL Batch & 400  & 14.56 & 0.97 & 12.9 & 28.43 \\
  & DL &  500 & 15.00 & 1.40  & 4.93 & 21.33 \\
  & DL batch  & 500 & 11.35 & 1.34   & 4.46 & 17.15 \\
\hline
 \multirow{4}{*}{200} & CL & 800 & 158.27 & 7.82 & 81.84 & 247.93\\
  & CL Batch &  800  & 133.78 & 7.82 & 81.58 & 223.18\\
  & DL   &  1000 & 126.61 & 9.9  & 21.45 & 157.96 \\
  & DL batch   &  1100 & 85.00 & 11.21   & 26.85 & 123.06 \\
\hline
 \multirow{4}{*}{220} & CL  &  1500  & 619.99 & 20.99 & 457.45 & 1098.43\\
  & CL Batch  &  1500 & 529.59 & 19.56 & 447.29 & 996.44\\
  & DL   & 1700 & 567.56 & 27.77   & 86.38 & 681.71 \\
  & DL batch   & 1600 & 540.37 & 24.23  & 73.76 & 638.36 \\
\hline
\end{tabular}
\end{center}
\end{table}

%The usual way to compute the regulator is to find vectors of the kernel of the relation matrix. However, the input of Maurer regulator computation function only needs power products corresponding to units. In \cite{BJregulator}, Biasse and Jacobson described a way of creating such an input involving system solving instead of kernel computation. It consists of creating extra relations $r_i$, $0\leq i\leq k$ for an integer $k$ large enough (in our experiments, we always have $k\leq 10$). Then, we solve the $k$ linear systems $X_iA = r_i$ using the function \texttt{certSolveRedLong} from the IML library \cite{iml}. Finally, we augment the matrix $A$ with the $k$ extra rows and the vectors $X_i$ with $k-1$ zero coefficients and a -1 coefficient at index $n+i$.
%\[ A' := \left( 
%   \begin{BMAT}(@)[2pt,1cm,1cm]{c}{c.c}
%   \begin{BMAT}(e){c}{c}
%A
%  \end{BMAT} \\
%\begin{BMAT}[2pt,1cm,0.5cm]{c}{c} 
%	r_i
%\end{BMAT}
%\end{BMAT}
%   \right)  \ \ 
%X_i' := \left(
%\begin{BMAT}(@)[2pt,3cm,0.5cm]{c.c}{c}
%\begin{BMAT}(e){c}{c}
%X_i
%  \end{BMAT} & 
%\begin{BMAT}[2pt,1cm,0.5cm]{ccc}{c}
%0\hdots 0 & -1 & 0\hdots 0
%\end{BMAT} 
%\end{BMAT}\right). 
%\]
%The $X'_i$ are kernel vectors of $A'$. In \cite{BJregulator}, it is shown that this method is faster than the one based on kernel computation because it only requires the solution to a few linear systems, and because it can be adapted in such a way that the linear system involves $A_{red}$. Indeed, we simply have to avoid eliminating the columns that appear in the $k$ extra rows. 

To solve the infrastructure discrete logarithm problem, we first need
to compute an approximation of the regulator.  For this purpose, we
used an improved version of Vollmer's system solving based algorithm
\cite{Vreg} described by Biasse and Jacobson \cite{BJregulator}.  In
order to find elements of the kernel, the algorithm creates extra
relations $r_i$, $0\leq i\leq k$ for some small integer $k$ (in our
experiments, we always have $k\leq 10$). Then, we solve the $k$ linear
systems $X_iA = r_i$ using the function \texttt{certSolveRedLong} from
the IML library \cite{iml}. We augment the matrix $A$ by adding the
$r_i$ as extra rows, and augment the vectors $X_i$ with $k-1$ zero
coefficients and a $-1$ coefficient at index $n+i,$ yielding
\[ A' := \left( 
   \begin{BMAT}(@)[2pt,1cm,1cm]{c}{c.c}
   \begin{BMAT}(e){c}{c}
A
  \end{BMAT} \\
\begin{BMAT}[2pt,1cm,0.5cm]{c}{c} 
	r_i
\end{BMAT}
\end{BMAT}
   \right), \ \ 
X_i' := \left(
\begin{BMAT}(@)[2pt,3cm,0.5cm]{c.c}{c}
\begin{BMAT}(e){c}{c}
X_i
  \end{BMAT} & 
\begin{BMAT}[2pt,1cm,0.5cm]{ccc}{c}
0 \,\hdots\, 0 & -1 & 0 \,\hdots\, 0
\end{BMAT} 
\end{BMAT}\right) \enspace. 
\]
The $X'_i$ are kernel vectors of $A',$ which can be used along with
the vector $\vec{v}$ containing the real parts of the relations, to
compute a multiple of the regulator with Maurer's algorithm
\cite[Sec~12.1]{MMaurer}.  As shown in Vollmer \cite{Vreg}, this
multiple is equal to the regulator with high probability.  In
\cite{BJregulator}, it is shown that this method is faster than the
one requiring a kernel basis because it only requires the solution to
a few linear systems, and it can be adapted in such a way that the
linear system involves $A_{red}$. 
%Indeed, we simply have to avoid
%eliminating the columns that appear in the $k$ extra rows.

Our algorithm to solve the infrastructure discrete logarithm problem
also makes use of the system solving algorithm. The input ideal
$\mathfrak{a}$ is first decomposed over the factor base, as in the
imaginary case, yielding the factorization $\mathfrak{a} = (\gamma)
\mathfrak{p_1}^{e_1} \cdots \mathfrak{p_n}^{e_n}.$ Then, we solve the
system $\vec{x} A = (e_1,\hdots,e_n)$ and compute a numerical
approximation to guaranteed precision of $\log |\alpha|$ modulo our
regulator multiple using Maurer's algorithm \cite[Sec~5.5]{MMaurer}
from $\gamma,$ the coefficients of $\vec{x},$ and the real parts of
the relation stored in $\vec{v}.$

The results of our experiments for the imaginary case are given in
Table~\ref{tab:imaginary}, and for the real case in
Table~\ref{tab:real}.  They were obtained on 2.4 GHz Xeon with 2GB of
memory.  For each bit length of $\Delta,$ denoted by
``$\mathrm{size}(\Delta)$,'' we list the average time in seconds
required to solve an instance of the appropriate discrete logarithm
problem ($\overline{t_\Delta})$ and standard deviation (std).  In the
imaginary case, for each discriminant size less than $220$ bits, $14$
instances of the discrete logarithm problem were solved. For size
$230$ and $256$ we solved $10,$ and for size $280$ and $300$ we solved
$5$ examples.  In the real case, $10$ instances were solved for each
size up to $256,$ $6$ for size $280,$ and $4$ for size $300.$
\begin{table}[!tp]
\caption{\label{tab:imaginary}Average run times for the discrete logarithm problem in
  $Cl_\Delta,$ $\Delta < 0$}
\begin{center}
\renewcommand{\arraystretch}{1.02}
\newdimen{\tempstrut}
\newsavebox\tempbox
\savebox{\tempbox}{$10^{8^f}$}
\tempstrut=\ht\tempbox
\begin{tabular}{|r||r|r|r|r|}
\hline
\multicolumn{1}{|>{\ }c<{\ }||}{$\mathrm{size}(\Delta)$} & 
\multicolumn{1}{c|}{$\overline{t_\Delta}$ (sec)} & 
\multicolumn{1}{c|}{std} &
\multicolumn{1}{>{\ }c<{\ }|}{$L_{|\Delta|}[1/2,\sqrt{2}] / \overline{t_\Delta}$} &
\multicolumn{1}{>{\ }c<{\ }|}{$L_{|\Delta|}[1/2,1] / \overline{t_\Delta}$} \\
\hline
140 &      7.89 &     2.33 & $ 6.44 \times 10^{8}$ & $1.79 \times 10^{8}\rule{0pt}{\tempstrut}$ \\
142 &      8.80 &     1.90 & $ 7.01 \times 10^{8}$ & $1.93 \times 10^{8}$ \\
144 &      9.91 &     3.13 & $ 7.55 \times 10^{8}$ & $2.06 \times 10^{8}$ \\
146 &     10.23 &     1.69 & $ 8.86 \times 10^{8}$ & $2.39 \times 10^{8}$ \\
148 &     11.80 &     3.45 & $ 9.29 \times 10^{8}$ & $2.48 \times 10^{8}$ \\
150 &     12.88 &     2.66 & $10.28 \times 10^{8}$ & $2.71 \times 10^{8}$ \\
152 &     14.42 &     3.38 & $11.09 \times 10^{8}$ & $2.89 \times 10^{8}$ \\
154 &     17.64 &     5.61 & $10.93 \times 10^{8}$ & $2.82 \times 10^{8}$ \\
156 &     22.06 &     5.57 & $10.53 \times 10^{8}$ & $2.69 \times 10^{8}$ \\
158 &     28.74 &    12.11 & $ 9.73 \times 10^{8}$ & $2.46 \times 10^{8}$ \\
160 &     27.12 &     8.77 & $12.39 \times 10^{8}$ & $3.10 \times 10^{8}$ \\
162 &     32.72 &    15.49 & $12.34 \times 10^{8}$ & $3.05 \times 10^{8}$ \\
164 &     31.08 &     6.85 & $15.58 \times 10^{8}$ & $3.82 \times 10^{8}$ \\
166 &     41.93 &    14.65 & $13.85 \times 10^{8}$ & $3.36 \times 10^{8}$ \\
168 &     51.92 &    16.51 & $13.39 \times 10^{8}$ & $3.21 \times 10^{8}$ \\
170 &     59.77 &    15.42 & $13.92 \times 10^{8}$ & $3.30 \times 10^{8}$ \\
172 &     68.39 &    17.79 & $14.54 \times 10^{8}$ & $3.42 \times 10^{8}$ \\
174 &     99.20 &    62.61 & $11.97 \times 10^{8}$ & $2.78 \times 10^{8}$ \\
176 &    124.86 &    80.29 & $11.35 \times 10^{8}$ & $2.61 \times 10^{8}$ \\
178 &    140.50 &    55.41 & $12.03 \times 10^{8}$ & $2.74 \times 10^{8}$ \\
180 &    202.42 &   145.98 & $ 9.94 \times 10^{8}$ & $2.24 \times 10^{8}$ \\
182 &    166.33 &    63.91 & $14.40 \times 10^{8}$ & $3.22 \times 10^{8}$ \\
184 &    150.76 &    58.37 & $18.90 \times 10^{8}$ & $4.18 \times 10^{8}$ \\
186 &    198.72 &    63.23 & $17.04 \times 10^{8}$ & $3.73 \times 10^{8}$ \\
188 &    225.90 &    94.94 & $17.79 \times 10^{8}$ & $3.86 \times 10^{8}$ \\
190 &    277.67 &   234.93 & $17.17 \times 10^{8}$ & $3.69 \times 10^{8}$ \\
192 &    348.88 &   134.36 & $16.20 \times 10^{8}$ & $3.45 \times 10^{8}$ \\
194 &    395.54 &   192.26 & $16.93 \times 10^{8}$ & $3.57 \times 10^{8}$ \\
196 &    547.33 &   272.83 & $14.48 \times 10^{8}$ & $3.02 \times 10^{8}$ \\
198 &    525.94 &   153.63 & $17.83 \times 10^{8}$ & $3.68 \times 10^{8}$ \\
200 &    565.43 &   182.75 &  $1.96 \times 10^{9}$ & $4.01 \times 10^{8}$ \\
202 &    561.36 &   202.80 &  $2.33 \times 10^{9}$ & $4.73 \times 10^{8}$ \\
204 &    535.29 &   205.68 &  $2.89 \times 10^{9}$ & $5.80 \times 10^{8}$ \\
206 &    776.64 &   243.35 &  $2.35 \times 10^{9}$ & $4.67 \times 10^{8}$ \\
208 &    677.43 &   200.08 &  $3.17 \times 10^{9}$ & $6.25 \times 10^{8}$ \\
210 &   1050.64 &   501.31 &  $2.41 \times 10^{9}$ & $4.70 \times 10^{8}$ \\
212 &   1189.71 &   410.98 &  $2.50 \times 10^{9}$ & $4.84 \times 10^{8}$ \\
214 &   1104.83 &   308.57 &  $3.17 \times 10^{9}$ & $6.07 \times 10^{8}$ \\
216 &   1417.64 &   352.27 &  $2.90 \times 10^{9}$ & $5.51 \times 10^{8}$ \\
218 &   2185.80 &   798.95 &  $2.21 \times 10^{9}$ & $4.16 \times 10^{8}$ \\
220 &   2559.79 &  1255.94 &  $2.22 \times 10^{9}$ & $4.13 \times 10^{8}$ \\
230 &   3424.40 &  1255.94 &  $3.66 \times 10^{9}$ & $6.52 \times 10^{8}$ \\
256 &  22992.70 & 13062.14 &  $4.00 \times 10^{9}$ & $6.36 \times 10^{8}$ \\
280 &  88031.08 & 34148.54 & $6.09 \times 10^{9}$ & $8.76 \times 10^{8}$ \\
300 &\ 702142.20 &\ 334566.51 & $3.16 \times 10^{9}$ & $4.19 \times 10^{8}$ \\
\hline
\end{tabular}
\end{center}
\end{table}

\begin{table}[!tp]
\caption{\label{tab:real}Average run times for the infrastructure discrete logarithm problem.}
\begin{center}
\renewcommand{\arraystretch}{1.02}
\savebox{\tempbox}{$10^{8^f}$}
\tempstrut=\ht\tempbox
\begin{tabular}{|r||r|r|r|r|}
\hline
\multicolumn{1}{|>{\ }c<{\ }||}{$\mathrm{size}(\Delta)$} & 
\multicolumn{1}{c|}{$\overline{t_\Delta}$ (sec)} & 
\multicolumn{1}{c|}{std} &
\multicolumn{1}{>{\ }c<{\ }|}{$L_{|\Delta|}[1/2,\sqrt{2}] /
  \overline{t_\Delta}$} &
\multicolumn{1}{>{\ }c<{\ }|}{$L_{|\Delta|}[1/2,1] / \overline{t_\Delta}$} \\
\hline
140 &     11.95 &     3.13 & $4.25 \times 10^{8}$ & $1.18 \times 10^{8}\rule{0pt}{\tempstrut}$ \\
142 &     12.47 &     2.06 & $4.95 \times 10^{8}$ & $1.36 \times 10^{8}$ \\
144 &     15.95 &     5.79 & $4.69 \times 10^{8}$ & $1.28 \times 10^{8}$ \\
146 &     14.61 &     2.94 & $6.20 \times 10^{8}$ & $1.67 \times 10^{8}$ \\
148 &     17.05 &     3.46 & $6.43 \times 10^{8}$ & $1.71 \times 10^{8}$ \\
150 &     21.65 &     4.55 & $6.12 \times 10^{8}$ & $1.61 \times 10^{8}$ \\
152 &     25.65 &     7.15 & $6.23 \times 10^{8}$ & $1.63 \times 10^{8}$ \\
154 &     29.01 &     6.97 & $6.65 \times 10^{8}$ & $1.72 \times 10^{8}$ \\
156 &     27.52 &     4.79 & $8.44 \times 10^{8}$ & $2.16 \times 10^{8}$ \\
158 &     33.59 &     8.80 & $8.32 \times 10^{8}$ & $2.10 \times 10^{8}$ \\
160 &     36.27 &    12.28 & $9.27 \times 10^{8}$ & $2.32 \times 10^{8}$ \\
162 &     43.55 &    10.73 & $9.27 \times 10^{8}$ & $2.29 \times 10^{8}$ \\
164 &     49.37 &    11.76 & $9.81 \times 10^{8}$ & $2.40 \times 10^{8}$ \\
166 &     59.73 &    17.18 & $9.72 \times 10^{8}$ & $2.36 \times 10^{8}$ \\
168 &     73.66 &    18.56 & $9.44 \times 10^{8}$ & $2.26 \times 10^{8}$ \\
170 &     75.50 &    19.80 & $1.10 \times 10^{9}$ & $2.62 \times 10^{8}$ \\
172 &    101.00 &    20.84 & $9.85 \times 10^{8}$ & $2.31 \times 10^{8}$ \\
174 &     94.80 &    38.87 & $1.25 \times 10^{9}$ & $2.91 \times 10^{8}$ \\
176 &    106.30 &    23.77 & $1.33 \times 10^{9}$ & $3.07 \times 10^{8}$ \\
178 &    149.70 &    44.04 & $1.13 \times 10^{9}$ & $2.57 \times 10^{8}$ \\
180 &    132.70 &    30.25 & $1.52 \times 10^{9}$ & $3.42 \times 10^{8}$ \\
182 &    178.80 &    25.67 & $1.34 \times 10^{9}$ & $2.99 \times 10^{8}$ \\
184 &    211.40 &    52.14 & $1.35 \times 10^{9}$ & $2.98 \times 10^{8}$ \\
186 &    258.20 &   110.95 & $1.31 \times 10^{9}$ & $2.87 \times 10^{8}$ \\
188 &    352.70 &    94.50 & $1.14 \times 10^{9}$ & $2.47 \times 10^{8}$ \\
190 &    290.90 &    46.57 & $1.64 \times 10^{9}$ & $3.52 \times 10^{8}$ \\
192 &    316.80 &    51.75 & $1.78 \times 10^{9}$ & $3.80 \times 10^{8}$ \\
194 &    412.90 &    71.90 & $1.62 \times 10^{9}$ & $3.42 \times 10^{8}$ \\
196 &    395.40 &    94.71 & $2.00 \times 10^{9}$ & $4.18 \times 10^{8}$ \\
198 &    492.30 &   156.69 & $1.90 \times 10^{9}$ & $3.94 \times 10^{8}$ \\
200 &    598.90 &   187.19 & $1.85 \times 10^{9}$ & $3.79 \times 10^{8}$ \\
202 &    791.40 &   285.74 & $1.65 \times 10^{9}$ & $3.35 \times 10^{8}$ \\
204 &    888.10 &   396.85 & $1.74 \times 10^{9}$ & $3.49 \times 10^{8}$ \\
206 &    928.40 &   311.37 & $1.96 \times 10^{9}$ & $3.90 \times 10^{8}$ \\
208 &   1036.10 &   260.82 & $2.07 \times 10^{9}$ & $4.08 \times 10^{8}$ \\
210 &   1262.30 &   415.32 & $2.00 \times 10^{9}$ & $3.91 \times 10^{8}$ \\
212 &   1582.30 &   377.22 & $1.88 \times 10^{9}$ & $3.64 \times 10^{8}$ \\
214 &   1545.10 &   432.42 & $2.27 \times 10^{9}$ & $4.34 \times 10^{8}$ \\
216 &   1450.80 &   453.85 & $2.84 \times 10^{9}$ & $5.39 \times 10^{8}$ \\
218 &   2105.00 &   650.64 & $2.30 \times 10^{9}$ & $4.32 \times 10^{8}$ \\
220 &   2435.70 &   802.57 & $2.33 \times 10^{9}$ & $4.34 \times 10^{8}$ \\
230 &   5680.90 &  1379.94 & $2.21 \times 10^{9}$ & $3.93 \times 10^{8}$ \\
256 &  29394.01 &  7824.15 & $3.13 \times 10^{9}$ & $4.98 \times 10^{8}$ \\
280 &  80962.80 & 27721.01 & $6.62 \times 10^{9}$ & $9.52 \times 10^{8}$ \\
300 &\ 442409.00 &\ 237989.12 & $5.01 \times 10^{9}$ & $6.64 \times 10^{8}$ \\
\hline
\end{tabular}
\end{center}
\end{table}

\clearpage
For the extrapolations in the next section, we need to have a good
estimate of the asymptotic running time of the algorithm.  As
described in the previous section, the best proven run time is
$O(L_{|\Delta|}[1/2,3 \sqrt{2}/4 + o(1)],$ but as we use sieving to
generate relations, this can likely be reduced to
$O(L_{|\Delta|}[1/2,1 + o(1)]).$ To test which running time is most
likely to hold for the algorithm we implemented, we list
$L_{|\Delta|}[1/2,3\sqrt{2}/4] / \overline{t_\Delta}$ and
$L_{|\Delta|}[1/2,1] / \overline{t_\Delta}$ in
Table~\ref{tab:imaginary} and Table~\ref{tab:real}.  In both cases,
our data supports the hypothesis that the run time of our algorithm is
indeed closer to $O(L_{|\Delta|}[1/2,1 + o(1)]),$ with the exception
of a few outliers corresponding to instances where only a few
instances of the discrete logarithm were computed for that size,

\section{Security Estimates}\label{sec:estimates}

General purpose recommendations for securely choosing discriminants
for use in quadratic field cryptography can be found in \cite{HMiq}
for the imaginary case and \cite{JSWkeyex} for the real case.  In both
cases, it usually suffices to use prime discriminants, as this forces
the class number $h_\Delta$ to be odd.  In the imaginary case, one
then relies on the Cohen-Lenstra heuristics \cite{CLheur} to guarantee
that the class number is not smooth with high probability.  In the
real case, one uses the Cohen-Lenstra heuristics to guarantee that the
class number is very small (and that the infrastructure is therefore
large) with high probability.

Our goal is to estimate what bit lengths of appropriately-chosen
discriminants, in both the imaginary and real cases, are required to
provide approximately the same level of security as the RSA moduli
recommended by NIST \cite{NIST-keys}.  The five security levels
recommended by NIST correspond to using secure block ciphers with keys
of $80,$ $112,$ $128,$ $192,$ and $256$ bits.  The estimates used by
NIST indicate that RSA moduli of size $1024,$ $2048,$ $3072,$ $7680,$
and $15360$ should be used.

To estimate the required sizes of discriminants, we follow the
approach of Hamdy and M{\"o}ller \cite{HMiq}, who provided such
estimates for the imaginary case.  Our results update these in the
sense that our estimates are based on our improved algorithms for
solving the discrete logarithms in quadratic fields, as well as the
latest data available for factoring large RSA moduli.  Our estimates
for real quadratic fields are the first such estimates produced.

Following, Hamdy and M{\"o}ller, suppose that an algorithm with
asymptotic running time $L_N[e,c]$ runs in time $t_1$ on input $N_1.$
Then, the running time $t_2$ of the algorithm on input $N_2$ can be
estimated using the equation
\begin{equation}\label{eq:estimate}
\frac{L_{N_1}[e,c]}{L_{N_2}[e,c]} = \frac{t_1}{t_2}\;.
\end{equation}
We can also use the equation to estimate an input $N_2$ that will
cause the algorithm to have running time $t_2,$ again given the time
$t_1$ for input $N_1.$

The first step is to estimate the time required to factor the RSA
numbers of the sizes recommended by NIST.  The best algorithm for
factoring large integers is the generalized number field sieve
\cite{LL93}, whose asymptotic running time is heuristically $L_N[1/3,
\sqrt[3]{64/9} + o(1)].$ To date, the largest RSA number factored is
RSA-768, a $768$ bit integer \cite{RSA768}.  It is estimated in
\cite{RSA768} that the total computation required $2000$ $2.2$ GHz AMD
Opteron years.  As our computations were performed on a different
architecture, we follow Hamdy and M{\"o}ller and use the MIPS-year
measurement to provide an architecture-neutral measurement.  In this
case, assuming that a $2.2$ GHz AMD Opteron runs at $4400$ MIPS, we
estimate that this computation took $8.8 \times 10^{6}$ MIPS-years.
Using this estimate in conjunction with \eqref{eq:estimate} yields the
estimated running times to factor RSA moduli of the sizes recommended
by NIST given in Table~\ref{tab:estimates}.  When using this method,
we use $N_1 = 2^{768}$ and $N_2 = 2^b,$ where $b$ is the bit length of
the RSA moduli for which we compute a run time estimate.

The second step is to estimate the discriminant sizes for which the
discrete logarithm problems require approximately the same running
time.  The results in Table~\ref{tab:imaginary} and
Table~\ref{tab:real} suggest that $L_N[1/2,1+o(1)]$ is a good estimate
of the asymptotic running time for both algorithms.  Thus, we use
$L_N[1/2,1]$ in \eqref{eq:estimate}, as ignoring the $o(1)$ results in
a conservative under-estimate of the actual running time.  For $N_1$
and $t_1,$ we take the largest discriminant size in each table for
which at least $10$ instances of the discrete logarithm problem were
run and the corresponding running time (in MIPS-years); thus we used
$256$ in the imaginary case and $230$ in the real case.  We take for
$t_2$ the target running time in MIPS-years.  To convert the times in
seconds from Table~\ref{tab:imaginary} and Table~\ref{tab:real} to
MIPS-years, we assume that the $2.4$ GHz Intel Xeon machine runs at
$4800$ MIPS.  To find the corresponding discriminant size, we simply
find the smallest integer $b$ for which $L_{2^b}[1/2,1] >
L_{N_1}[1/2,1] t_2 / t_1.$

Our results are listed in Table~\ref{tab:estimates}.  We list the size
in bits of RSA moduli (denoted by ``$RSA$''), discriminants of
imaginary quadratic fields (denoted by ``$\Delta$ (imaginary)''), and
real quadratic fields (denoted by ``$\Delta$ (real'') for which
factoring and the quadratic field discrete logarithm problems all have
the same estimated running time.  For comparison purposes, we also
list the discriminant sizes recommended in \cite{HMiq}, denoted by
``$\Delta$ (imaginary, old).''  Note that these estimates were based
on different equivalent MIPS-years running times, as the largest
factoring effort at the time was RSA-512.  In addition, they are based
on an implementation of the imaginary quadratic field discrete
logarithm algorithm from \cite{Jdl}, which is slower than the improved
version from this paper.  Consequently, our security parameter
estimates are slightly larger than those from \cite{HMiq}.  We note
also that the recommended discriminant sizes are slightly smaller in
the real case, as the infrastructure discrete logarithm problem
requires more time to solve on average than the discrete logarithm in
the imaginary case.
\begin{table}[!ht]
\caption{Security Parameter Estimates}\label{tab:estimates}
\begin{center}
\renewcommand{\arraystretch}{1.02}
\savebox{\tempbox}{$10^{8^f}$}
\tempstrut=\ht\tempbox
\begin{tabular}{|r||r|r|r|r|}
\hline
\multicolumn{1}{|>{\ }c<{\ }||}{RSA} & 
\multicolumn{1}{>{\ }c<{\ }|}{$\Delta$ (imaginary, old)} & 
\multicolumn{1}{>{\ }c<{\ }|}{$\Delta$ (imaginary)} & 
\multicolumn{1}{>{\ }c<{\ }|}{$\Delta$ (real)} &
\multicolumn{1}{>{\ }c<{\ }|}{Est.\ run time (MIPS-years)} \\
\hline
  768 &  540 &  640 &  634 & $8.80 \times 10^{6}\rule{0pt}{\tempstrut}$ \\
 1024 &  687 &  798 &  792 & $1.07 \times 10^{10}$ \\
 2048 & 1208 & 1348 & 1341 & $1.25 \times 10^{19}$ \\
 3072 & 1665 & 1827 & 1818 & $4.74 \times 10^{25}$ \\
 7680 &    0 & 3598 & 3586 & $1.06 \times 10^{45}$ \\
15360 &    0 & 5971 & 5957 & $1.01 \times 10^{65}$ \\
\hline
\end{tabular}
\end{center}
\end{table}

\section{Conclusions}

It is possible to produce more accurate security parameter estimates
by taking more factors into account as is done, for example, by
Lenstra and Verheul \cite{LVselect}, as well as using a more accurate
performance measure than MIPS-year.  However, our results nevertheless
provide a good rough guideline on the required discriminant sizes that
is likely sufficiently accurate in the inexact science of predicting
security levels.

It would also be of interest to conduct a new comparison of the
efficiency of RSA as compared to the cryptosystems based on quadratic
fields.  Due to the differences in the asymptotic complexities of
integer factorization and the discrete logarithm problems in quadratic
fields, it is clear that there is a point where the cryptosystems
based on quadratic fields will be faster than RSA.  However, ideal
arithmetic is somewhat more complicated than the simple integer
arithmetic required for RSA, and in fact Hamdy's conclusion
\cite{Hamdy} was that even with smaller parameters, cryptography using
quadratic fields was not competitive at the security levels of
interest.  There have been a number of recent advances in ideal
arithmetic in both the imaginary and real cases (see, for example,
\cite{IJScubing} and \cite{JSWnucomp}) that warrant revisiting this
issue.

%
% ---- Bibliography ----
%
\bibliographystyle{amsplain}

\begin{thebibliography}{10}

\bibitem{BBounds}
E.~Bach, \emph{Explicit bounds for primality testing and related problems},
  Math. Comp. \textbf{55} (1990), no.~191, 355--380.

\bibitem{bernstein}
D.~Bernstein, \emph{How to find smooth parts of integers}, submitted to
  \textit{Mathematics of Computation}.

\bibitem{biasse}
J.-F. Biasse, \emph{Improvements in the computation of ideal class groups of
  imaginary quadratic number fields}, To appear in \textit{Advances in
  Mathematics of Communications}, see
  \url{http://www.lix.polytechnique.fr/~biasse/papers/biasseCHILE.pdf}.

\bibitem{BJregulator}
J.-F. Biasse and M.~J. Jacobson, Jr., \emph{Practical improvements to class
  group and regulator computation of real quadratic fields}, 2010, To appear in
  ANTS 9.

\bibitem{BSub}
J.~Buchmann, \emph{A subexponential algorithm for the determination of class
  groups and regulators of algebraic number fields}, S\'{e}minaire de
  Th\'{e}orie des Nombres (Paris), 1988--89, pp.~27--41.

\bibitem{BVforms}
J.~Buchmann and U.~Vollmer, \emph{Binary quadratic forms: An algorithmic
  approach}, Algorithms and Computation in Mathematics, vol.~20,
  Springer-Verlag, Berlin, 2007.

\bibitem{BWKeyEx}
J.~Buchmann and H.~C. Williams, \emph{A key-exchange system based on imaginary
  quadratic fields}, Journal of Cryptology \textbf{1} (1988), 107--118.

\bibitem{BWKeyExReal}
\bysame, \emph{A key-exchange system based on real quadratic fields}, CRYPTO
  '89, Lecture Notes in Computer Science, vol. 435, 1989, pp.~335--343.

\bibitem{cavallar}
S.~Cavallar, \emph{Strategies in filtering in the number field sieve}, ANTS-IV:
  Proceedings of the 4th International Symposium on Algorithmic Number Theory,
  Lecture Notes in Computer Science, vol. 1838, Springer-Verlag, 2000,
  pp.~209--232.

\bibitem{iml}
Z.~Chen, A.~Storjohann, and C.~Fletcher, \emph{{IML: Integer Matrix Library}},
  available at \url{http://www.cs.uwaterloo.ca/~z4chen/iml.html}, 2007.

\bibitem{CLheur}
H.~Cohen and H.~W. Lenstra, Jr., \emph{Heuristics on class groups of number
  fields}, Number Theory, Lecture Notes in Math., vol. 1068, Springer-Verlag,
  New York, 1983, pp.~33--62.

\bibitem{HMSub}
J.~L. Hafner and K.~S. McCurley, \emph{A rigorous subexponential algorithm for
  computation of class groups}, J. Amer. Math. Soc. \textbf{2} (1989),
  837--850.

\bibitem{Hamdy}
S.~Hamdy, \emph{{\"U}ber die {S}icherheit und {E}ffizienz kryptografischer
  {V}erfahren mit {K}lassengruppen imagin{\"a}r-quadratischer
  {Z}ahlk{\"o}rper}, Ph.D. thesis, Technische Universit{\"a}t Darmstadt,
  Darmstadt, Germany, 2002.

\bibitem{HMiq}
S.~Hamdy and B.~M{\"o}ller, \emph{Security of cryptosystems based on class
  groups of imaginary quadratic orders}, Advances in Cryptology - ASIACRYPT
  2000, Lecture Notes in Computer Science, vol. 1976, 2000, pp.~234--247.

\bibitem{HJWnonint_full}
D.~H{\"u}hnlein, M.~J. Jacobson, Jr., and D.~Weber, \emph{Towards practical
  non-interactive public-key cryptosystems using non-maximal imaginary
  quadratic orders}, Designs, Codes and Cryptography \textbf{30} (2003), no.~3,
  281--299.

\bibitem{IJScubing}
L.~Imbert, M.~J. Jacobson, Jr., and A.~Schmidt, \emph{Fast ideal cubing in
  imaginary quadratic number and function fields}, To appear in to Advances in
  Mathematics of Communication, 2010.

\bibitem{Jdl}
M.~J. Jacobson, Jr., \emph{Computing discrete logarithms in quadratic orders},
  Journal of Cryptology \textbf{13} (2000), 473--492.

\bibitem{JSWkeyex}
M.~J. Jacobson, Jr., R.~Scheidler, and H.~C. Williams, \emph{The efficiency and
  security of a real quadratic field based key exchange protocol}, Public-Key
  Cryptography and Computational Number Theory (Warsaw, Poland), de Gruyter,
  2001, pp.~89--112.

\bibitem{JSWnucomp}
\bysame, \emph{An improved real quadratic field based key exchange procedure},
  Journal of Cryptology \textbf{19} (2006), 211--239.

\bibitem{JWPellBook}
M.~J. Jacobson, Jr. and H.~C. Williams, \emph{Solving the {P}ell equation}, CMS
  Books in Mathematics, Springer-Verlag, 2009, ISBN 978-0-387-84922-5.

\bibitem{RSA768}
T.~Kleinjung, K.~Aoki, J.~Franke, A.~K. Lenstra, E.~Thom{\'e}, J.~W. Bos,
  P.~Gaudry, A.~Kruppa, P.~L. Montgomery, D.~A. Osvik, H.~te~Riele,
  A.~Timofeev, and P.~Zimmerman, \emph{Factorization of a 768-bit {RSA}
  modulus}, Eprint archive no.~2010/006, 2010.

\bibitem{LL93}
A.~K. Lenstra and H.~W. Lenstra, Jr., \emph{The development of the number field
  sieve}, Lecture Notes in Mathematics, vol. 1554, Springer-Verlag, Berlin,
  1993.

\bibitem{LVselect}
A.~K. Lenstra and E.~Verheul, \emph{Selecting cryptographic key sizes},
  Proceedings of Public Key Cryptography 2000, Lecture Notes in Computer
  Science, vol. 1751, 2000, pp.~446--465.

\bibitem{MMaurer}
M.~Maurer, \emph{Regulator approximation and fundamental unit computation for
  real-quadratic orders}, Ph.D. thesis, Technische Universit{\"a}t Darmstadt,
  Darmstadt, Germany, 2000.

\bibitem{NIST-keys}
{National Institute of Standards and Technology (NIST)}, \emph{Recommendation
  for {K}ey {M}anagement --- {P}art 1: {G}eneral ({R}evised)}, NIST Special
  Publication 800-57, March, 2007, see:
  \url{http://csrc.nist.gov/groups/ST/toolkit/documents/SP800-57Part1_3-8-07.pdf}.

\bibitem{Vdl}
U.~Vollmer, \emph{Asymptotically fast discrete logarithms in quadratic number
  fields}, Algorithmic Number Theory --- ANTS-IV, Lecture Notes in Computer
  Science, vol. 1838, 2000, pp.~581--594.

\bibitem{Vreg}
\bysame, \emph{An accelerated {B}uchmann algorithm for regulator computation in
  real quadratic fields}, Algorithmic Number Theory --- ANTS-V, Lecture Notes
  in Computer Science, vol. 2369, 2002, pp.~148--162.

\bibitem{vollmer-thesis:2003}
\bysame, \emph{Rigorously analyzed algorithms for the discrete logarithm
  problem in quadratic number fields}, Ph.D. thesis, Technische
  Universit{\"{a}}t Darmstadt, 2003.

\end{thebibliography}

\providecommand{\bysame}{\leavevmode\hbox to3em{\hrulefill}\thinspace}
\providecommand{\MR}{\relax\ifhmode\unskip\space\fi MR }
% \MRhref is called by the amsart/book/proc definition of \MR.
\providecommand{\MRhref}[2]{%
  \href{http://www.ams.org/mathscinet-getitem?mr=#1}{#2}
}
\providecommand{\href}[2]{#2}

\end{document}